\documentclass[a4paper]{article}
\usepackage[latin1]{inputenc}
\sf
\usepackage{amssymb}

\usepackage{amssymb}

\newtheorem{lem}{Lemma}
\begin{document}
\title{Erratum}     
\author{Friedrich Wagemann \\
        Institut Girard Desargues -- ESA 5028 du CNRS\\
	Universit\'e Claude Bernard Lyon-I\\
        43, bd du 11 Novembre 1918 \\
	69622 Villeurbanne Cedex FRANCE\\
        tel.: +33.4.72.43.11.90\\
        fax:  +33.4.72.43.00.35\\
        e-mail: wagemann@desargues.univ-lyon1.fr}
\maketitle

F. Wagemann: Some Remarks on the Cohomology of Krichever-Novikov
Algebras, {\it Lett. Math. Phys.} {\bf 47 } (1999),  173--177\\

There is a mistake in the statement of the main lemma. It should
read: 

\begin{lem}
Let $\Sigma_r=\Sigma\setminus\{p_1,\ldots,p_r\}$ be an open Riemann
surface of finite type. 

Suppose that $\Sigma_r$ is trivializable, i.e. that the algebraic
tangent bundle of the corresponding affine cuve is trivial. Then
$\overline{Mer(\Sigma_r)} = Hol(\Sigma_r)$,

in full words,

in the topology of uniform convergence on the compact sets $K_n$,
a function with essential singularities in the points $p_k$
may be approximated by functions with poles in the points.
\end{lem}

The proof given in the article obviously works only for this case, for
an open Riemann surface is always holomorphically trivializable, but
the globally non-zero vector field is meromorphic only under the above
hypothesis. 

The trivializability condition can be expressed as the hypothesis that
the divisor associated to the tangent bundle of the affine algebraic
curve lies in the linear span of the points $p_1,\ldots,p_r$. This is
only valid in very rare cases, the points $p_1,\ldots,p_r$ being
required to lay at special places, except for genus 0 and 1 where it
is always true ($r\geq 1$). 

The question whether this lemma remains true even for
non-trivializable Riemann surfaces of finite type is - to our
knowledge - open.

{\bf Acknowledgements:}

The author thanks A. Rieman for very useful discussions on the subject.

\end{document}